\documentstyle[12pt]{article}
\begin{document}
\begin{center}
{\large \bf 
Zipf's Law Distributions for Korean Stock Prices \\

\vspace*{.5in}

\normalsize 
Kyungsik Kim$^{*}$, S. M. Yoon$^{a}$, C. Christopher Lee$^{b}$ and K. H. Chang$^{c}$

\vspace*{.3in}

{\em 
Department of Physics, Pukyong National University,\\
Pusan 608-737, Korea \\
$^{a}$Division of Economics, Pukyong National University,\\
Pusan 608-737, Korea \\ 
$^{b}$Department of Business Administration, Central Washington University,\\
WA 98926, USA \\
$^{c}$ Forecast Research Laboratory, Meteorological Research \\ 
Institute, KMA, Seoul 156-720, South Korea } \\
}
\end{center}

\hfill\\
%
%
%
\baselineskip 24pt

This paper investigates the rank distribution, cumulative probability, and 
probability density of price returns for the stocks traded in the KSE and the KOSDAQ market. 
This research demonstrates that the rank distribution is consistent approximately
with the Zipf's law with exponent $\alpha = -1.00$ (KSE) and $-1.31$ (KOSDAQ), 
similar to that of stock prices traded on the TSE. In addition, the cumulative probability 
distribution follows a power law with scaling exponent $\beta = -1.23$ (KSE) and $-1.45$ (KOSDAQ). 
In particular, the evidence displays that the probability density of normalized price returns 
for two kinds of assets almost has the form of an exponential function, 
similar to the result in the TSE and the NYSE.

\vskip 10mm 
\hfill\\
PACS numbers: 89.65.Gh, 02.50.Ey, 05.40.Jc, 05.45.Tp \\
Keywords: Zipf's law; Power law; KOSPI; KOSDAQ  \\
%
\hfill\\
$^{*}$E-mail address: kskim@pknu.ac.kr;
Tel.: +82-51-620-6354; Fax: +82-51-611-6357 \\   

\newpage


At the end of nineteenth century, Pareto$^{1)}$ studied the income and wealth distributions 
characteristic of national economies. Gini$^{2)}$ showed the income distribution approximated 
by a power law with a non-universal exponent, while Gibrat$^{3)}$ reported a log-normal distribution 
of income based on multiplicative random processes. Particularly, Mandelbrot$^{4)}$ discovered 
the distribution of incomes as a power law, and Montroll and Schlesinger$^{5)}$
confirmed that the income distribution of a high-income group involves a power law, 
while that of a low-income group follows a log-normal distribution. Recently, 
a distribution of American company sizes has been found to be closer to the log-normal law.$^{6)}$
The databases of Japanese and international companies, analyzed and reported by Okuyama and Takayasu,$^{7)}$
suggest that the distribution function of the annual income of companies involves a power law distribution 
consistent with the Zipf's law. Furthermore, many researchers have attempted to study and calculate 
the feature of distribution functions characterized by the Zipf's law, in scientific fields such as voting processes,$^{8)}$
firm bankruptcy,$^{9)}$ investment strategy,$^{10)}$ expressed genes,$^{11)}$ and earthquakes.$^{12)}$

In the view of this article, it would be of high importance to elucidate the scaling properties and
statistical methods for stock prices. Thus, the focus of this paper will be upon discovering 
the scaling relation for Korea stock prices traded on the Korea Stock Exchange (KSE) and 
the Korean Securities Dealers Automated Quotations (KOSDAQ) markets. KSE is a major market place 
of common stocks for large Korean firms, while KOSDAQ is the over-the-counter marketplace under 
the auspices of the Korea Securities Dealers Association and is tasked with corporate financing 
for knowledge-based ventures, high-tech companies and small to medium sized enterprises. 
This paper mainly calculates the rank distribution, cumulative probability, 
and probability density of price returns of the above two markets.

First, this study analyzes the database of listed companies for two assets in Korean stock market. 
The rank for asset $R(p)$ is represented in terms of:  
\begin{equation}
\log R (p ) =  \log b + \alpha \log p ,
\label{eq:a1}
\end{equation}
where $ \alpha $ and $b$ denote, respectively, the scaling exponent and the parameter, 
and $R(p)$ denotes the rank of companies for stock prices traded on the KSE and the KOSDAQ. 
Eq. $(1)$ is satisfied with the Zipf's law in the case of $ \alpha = -1$. This paper collected data 
from $668$ Korean companies listed in the KSE, and $620$ firms listed in the KOSDAQ, 
for one-day period on December $27$, $2002$. The rank distributions follows the Zipf's law 
with exponent $ \alpha = -1.00$ (KSE in Fig.1) and $-1.31$ (KOSDAQ). Results are consistent 
approximately with that of stock prices traded on the Tokyo Stock Exchange (TSE) in Japan, 
but different from the estimated rank of stock prices on the New York Stock Exchange (NYSE) 
in the U.S. characterized by the power law.$^{13)}$

Next, the cumulative probability of a stock price is defined as: 
\begin{equation}
  P(\geq p) \propto p^{\beta},
\label{eq:b2}
\end{equation}
where $ \beta $ is the scaling exponent for the cumulative probability of stock prices. 
From our calculated result, cumulative probability distributions follow the power law approximately with scaling exponents 
$ \beta = -1.23$ (KSE) and $-1.45$ (KOSDAQ in Fig. $2$), similar to the findings from the TSE.$^{13)}$
For income distributions, it is widely accepted that the Japanese income distribution$^{7)}$
is closer to Zipf's law, while the power law holds in the U.S. and Korean income distributions.$^{6,14)}$

Lastly let $p(t)$ be a stock price at time $t$. The price return, i.e. the ratio of the stock price of successive times,
is defined by $ r(t)= \log p(t+ \tau )/p(t)$, where $ \tau $ is the time lag. We introduce 
the probability density of price returns as a function of the normalized price return
$ r^{*} =(r- <r>)/ \sigma $, where $<r>$ and $ \sigma $ denote, respectively, the mean value and 
the standard deviation of $r(t)$. It is then found that the probability density of price returns 
almost has the form of an exponential function as
\begin{equation}
P(r^{*} )  \propto \exp (\kappa r^{*} ),
\label{eq:c3}
\end{equation}
where $ \kappa $ is a proportional coefficient. Fig. $3$ depicts the probability density of price returns 
as a function of the normalized price return with $ \kappa $ = $-0.56$ (KSE), similar to the findings 
from the KOSDAQ ($ \kappa $ = $-0.72$), TSE, and NYSE.$^{13)}$
 
In summary, this research establishes that the rank distribution function, 
the cumulative probability, and the probability density of price returns from stock prices 
traded on two stock markets - the KSE and the KOSDAQ in Korea. As well, 
the results of this study are consistent with earlier works that suggest many 
international assets are characterized by Zipf's law or power law. Particularly, 
it is found that the rank for stock prices traded on the KSE, KOSDAQ and TSE follows 
the the Zipf's law, while that of the NYSE displays the power law. The cumulative probability 
of the NYSE is consistent with the power law, significantly different from Zipf's law in the KSE, 
KOSDAQ, and TSE. It is also revealed that the probability density of normalized price returns 
almost behaves as an exponential function. This sort of research would be of great value in
characterizing and categorizing the scaling relations in the Zipf's and Pareto's laws 
in other foreign financial markets. In a projected work, the intention will be to investigate
the tick data of other stock prices, and compare these with calculations performed in other nations.

This work was supported by Korea Research Foundation Grant(KRF-$2003$-$002$-B$00032$). \\

\newpage
\vskip 10mm
\begin{center}
{\bf FIGURE  CAPTIONS}
\end{center}

\vspace {5mm}

\noindent
Fig. $1$.  Distribution of the rank for stocks traded on the KSE, where the least-squares fit suggests 
the power law with exponent $ \alpha = -1.00$.  The solid line shows the Gaussian distribution
$A \exp (-cp^{2} )$ with $A=560$ and $c=4.0 \times 10^{-8}$.
\vspace {10mm}

\noindent
Fig. $2$.  Plot of cumulative probability of stocks traded on the KOSDAQ, where the slope of the dot line 
is $ \beta = -1.45$.

\vspace {10mm}

\noindent
Fig. $3$.  Probability density of price returns for stocks traded on the KSE as a function of 
the normalized price return with $ \kappa = -0.56$. The solid line shows the Gaussian distribution 
$ B \exp (- b r^{*} )$ with $ B=0.3$ and $ \sigma=1$.
\vspace {10mm}

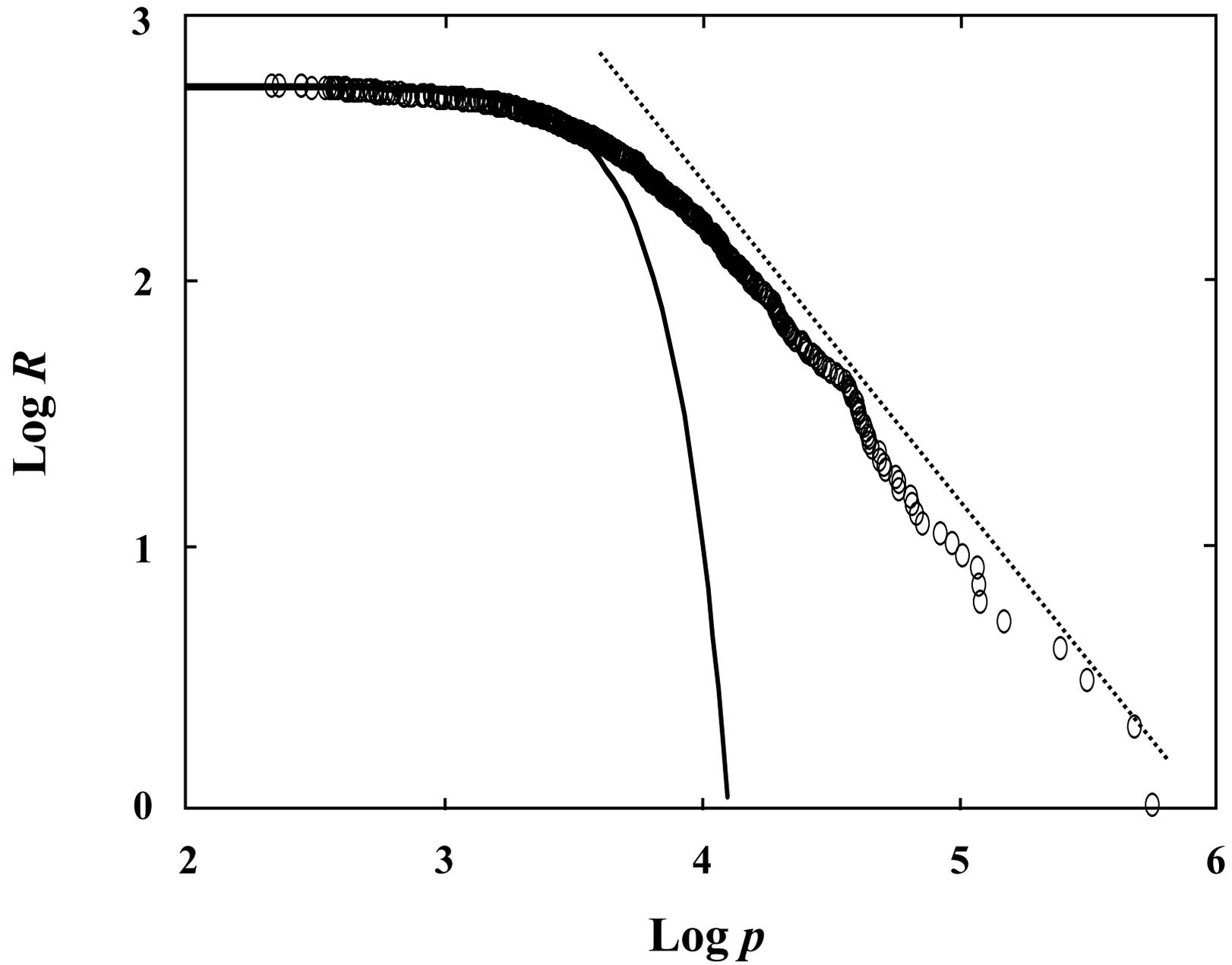
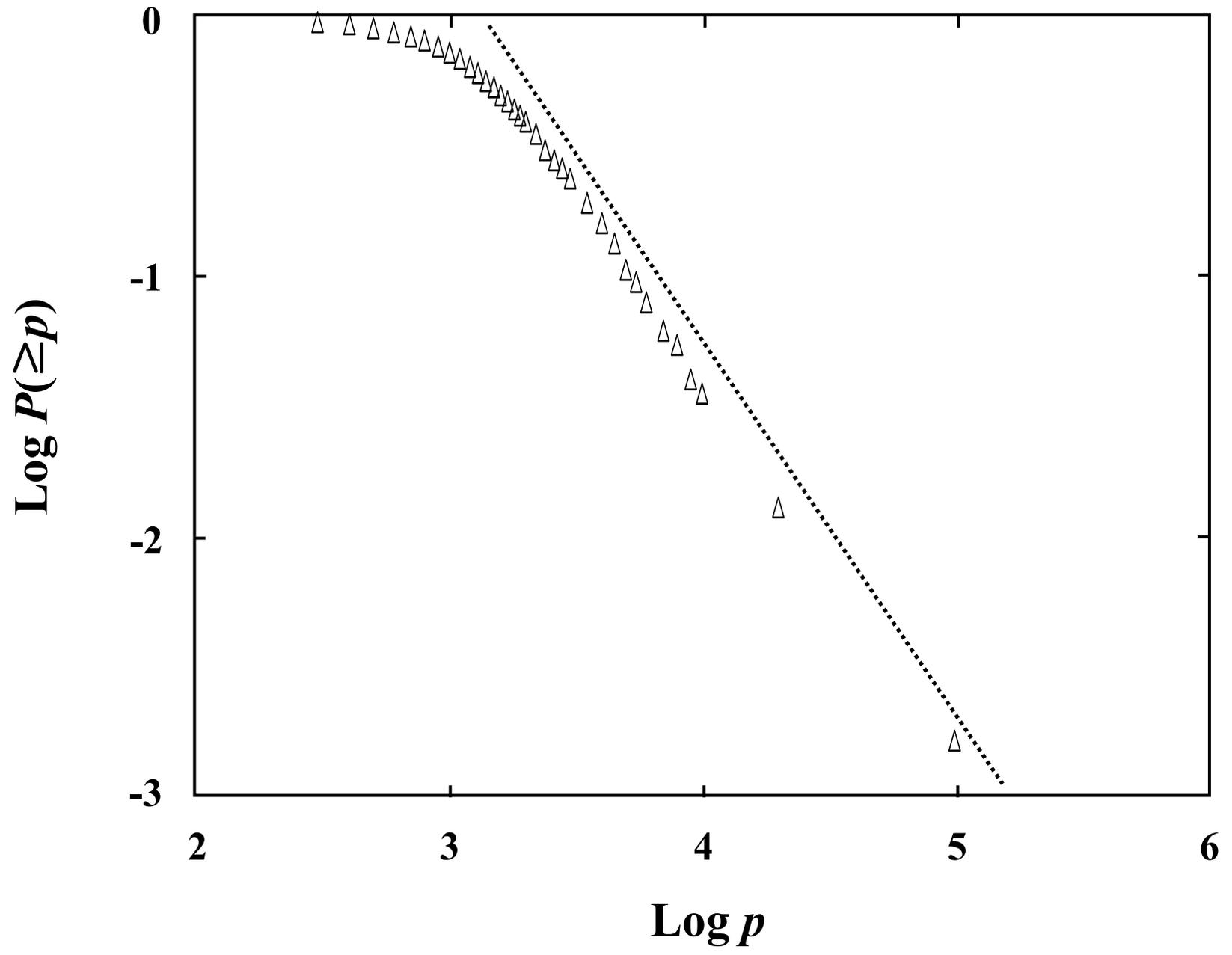
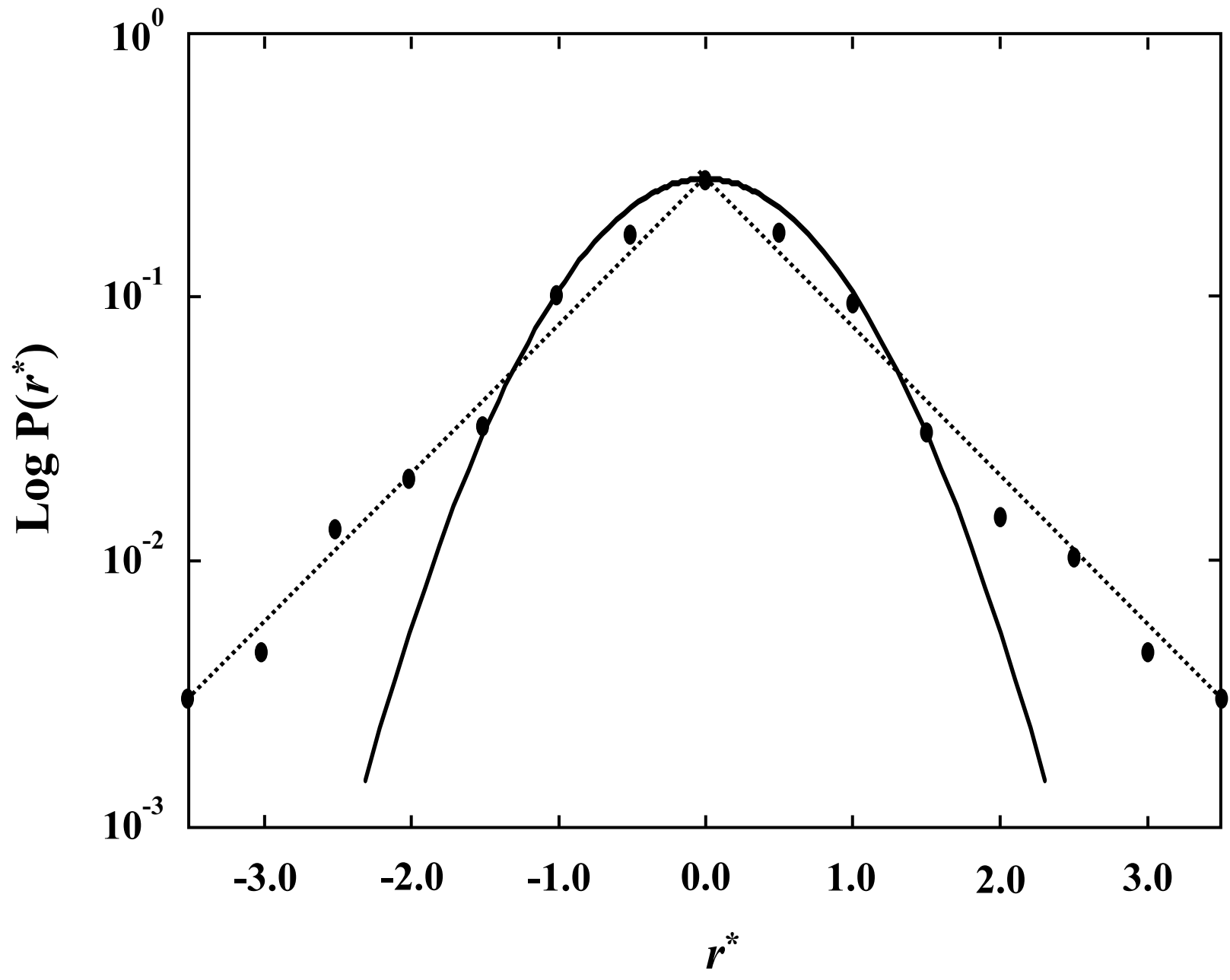
\end{document}